\documentstyle[prl,aps]{revtex}
\input epsf

\begin{document}

\twocolumn

\title{Observation of p-wave Threshold Law Using Evaporatively Cooled Fermionic Atoms}

\author{B. DeMarco,
J. L. Bohn, J. P. Burke, Jr., M. Holland, and D. S. Jin
\cite{adr1}}
\address{JILA, \\ National Institute of Standards and Technology
and University of Colorado,
\\ and \\ Physics Department, University of Colorado, Boulder, CO
80309-0440} \date{\today} \maketitle

\begin{abstract}
We have measured independently both s-wave and p-wave
cross-dimensional thermalization rates for ultracold $^{40}K$
atoms held in a magnetic trap.  These measurements reveal that
this fermionic isotope has a large positive s-wave triplet
scattering length in addition to a low temperature p-wave shape
resonance. We have observed directly the p-wave threshold law
which, combined with the Fermi statistics, dramatically suppresses
elastic collision rates at low temperatures.  In addition, we
present initial evaporative cooling results that make possible
these collision measurements and are a precursor to achieving
quantum degeneracy in this neutral, low-density Fermi system.

\end{abstract} 

\narrowtext

While many examples of quantum degenerate fermionic systems are
found in nature (electrons in metals and nucleons in nuclear
matter, for example), low density Fermi systems are exceedingly
rare. However, techniques similar to those that led to the
observation of Bose-Einstein condensation (BEC) in atomic systems
\cite{BEC} can be exploited to realize a quantum degenerate,
dilute gas of fermionic atoms.  Novel phenomena predicted for this
system include the suppression of inelastic collisions
\cite{noinelastic}, linewidth narrowing through suppression of
spontaneous emission \cite{narrowline}, and the possibility of a
phase transition to a superfluid-like state at sufficiently low
temperatures \cite{FDBEC}.  Just as was found in the case of BEC
in alkali atoms, knowledge of the binary elastic collision
cross-sections and interatomic potentials is crucial for
experiments on fermionic species.  Both evaporative cooling and
the prospect of fermionic superfluidity depend on the cold
collision parameters.  For example, accurate prediction of
magnetic-field Feshbach resonances \cite{Feshbach}, which could be
used to realize Cooper pairing of fermionic atoms, hinges on
detailed understanding of the interatomic potentials. In this
letter, we present measurements of elastic collision cross
sections for evaporatively cooled $^{40}$K, including a direct
observation of p-wave threshold behavior and the resultant strong
suppression of the collision rate.

Among the stable fermionic alkali atoms, $^{40}$K yields the
greatest range of possibilities for evaporative cooling strategies
and interaction studies because of its large atomic spin, F (F=9/2
and F=7/2 hyperfine ground states). In addition to having a large
number of spin states that can be held in the usual magnetic
traps, potassium has two bosonic isotopes, $^{39}$K and $^{41}$K,
which could be used for future studies of mixed boson-fermion
dilute gases.  Forced evaporative cooling \cite{evap}, which has
proven essential for achieving quantum degeneracy in bosonic
alkali gases, relies on elastic collisions for rethermalization of
the trapped atomic gas. However, evaporative cooling strategies
for fermionic samples are complicated by the fact that atomic
collisions at these low temperatures (100's of $\mu$K and below)
are predominantly s-wave in character for bosonic atoms, while the
Pauli exclusion principle prohibits s-wave collisions between
spin-polarized fermions. Evaporative cooling for fermionic atoms
must then proceed either through p-wave collisions
\cite{pwavecooling} or through sympathetic cooling \cite{Myatt}.

Since the s-wave and p-wave binary elastic cross sections are not
well known for $^{40}$K \cite{Verhaar,Cote,Burke}, we have made
measurements of elastic collision rates in a magnetic trap.  The
Fermi-Dirac quantum statistics of $^{40}$K provide a unique
opportunity to observe p-wave collisions directly.  Exploiting
this fact, we have seen evidence for a p-wave shape resonance and
have observed threshold behavior of the p-wave cross section.  To
our knowledge, this measurement represents the first direct
verification of the p-wave threshold law for neutral scatterers.

We use a double-MOT (magneto-optical trap) apparatus
\cite{doubleMOT} to trap and pre-cool $^{40}$K atoms prior to
loading them into a purely magnetic trap. Operation of the MOT's
employs two MOPA (master-oscillator power-amplifier) diode laser
systems \cite{SDL}, each frequency stabilized using the DAVLL
(dichroic absorption vapor laser lock) technique \cite{davll}
which accomplishes a large frequency range for locking.  Also, we
have developed an atom source \cite{RSI} that is enriched with
5$\%$ $^{40}$K (whose natural abundance is 0.01$\%$).  This system
allows us to trap any of the three potassium isotopes, and to trap
$10^8$ $^{40}$K atoms (four orders of magnitude more $^{40}$K
atoms than previous efforts \cite{italians}).

Immediately before loading the sample into the magnetic trap, the
atoms are cooled to approximately 150 $\mu$K during a Doppler
cooling stage of the MOT in which the trapping light is jumped
closer to resonance.  Further, an optical pumping pulse transfers
the majority of the atoms into magnetically trappable states in
the F=9/2 hyperfine ground state.  The atoms are then loaded into
a cloverleaf magnetic trap \cite{cloverleaf} and after an initial
evaporative cooling stage, we are left with roughly $10^7$
$^{40}$K atoms at 60 $\mu$K and a peak density of $10^9$
cm$^{-3}$. The cloverleaf magnetic trap provides a cylindrically
symmetric harmonic potential, with a characteristic radial
frequency $\nu_r=44\pm1$ and an axial frequency $\nu_z$ of
$19\pm1$ Hz for loading.  The lifetime of the atoms in the
magnetic trap, limited by collisions with room-temperature
background atoms, has an exponential time constant of $300\pm50$
s, giving ample time for thermal relaxation studies as well as for
evaporation. The radial frequency and the bias magnetic field can
be altered smoothly by changing the current through a pair of
Helmholtz bias coils.

We determine elastic collision cross sections from measurements of
cross-dimensional thermalization rates \cite{elastic} in the
magnetic trap.  The sample is taken out of thermal equilibrium by
changing $\nu_r$ through a ramp of the bias coil current. For the
measurements reported here $\nu_r$ lies between 44 and 133 Hz. The
change in $\nu_r$ occurs adiabatically (slow compared to the
atomic motion in the trap) but much faster then the rate of
collisions between atoms. Since the axial frequency is essentially
unchanged, energy is added to (or removed from) the cloud in only
the radial dimension.  Elastic collisions then move energy between
the radial and the axial dimensions, and the thermal relaxation is
observed by monitoring the time evolution of the cloud's aspect
ratio.

To avoid perturbations to the image due to the spatially dependent
magnetic fields, the trap is turned off suddenly and the cloud is
imaged after 2.7 ms of free expansion. The aspect ratio of the
cloud is observed via absorption imaging using a 9.1 $\mu$s pulse
of light resonant with the 4S$_{1/2}$, F=9/2 to 4P$_{3/2}$, F=11/2
transition.  Optical depth is calculated from the image captured
on a CCD array and then surface fit to a gaussian distribution to
find the rms cloud size in both the radial and axial dimensions.
An example of the cloud evolution following a change in trap
potential is shown in Fig. 1.  Since the expanded cloud sizes are
proportional to the square root of the cloud energy in each
dimension, the exponential time constant for the redistribution of
energy, $\tau$, can be extracted from an appropriate fit to the
aspect ratio vs time.  To rule out significant relaxation through
trap anharmonicities, we have verified that the relaxation rate
$1/\tau$ scales linearly with the number of trapped atoms $N$.

To obtain the elastic collision cross section $\sigma$ from our
measurements of thermal relaxation rates, we use the relation:
$1/\tau={2 \over \alpha} n\sigma v$, where $n$ is the
density-weighted density of the trapped atoms given by ${1\over N}
\int n(r)^2d^3r$, $v$ is the rms relative velocity between two
atoms in the trap, and $\alpha$ is the calculated average number
of binary collisions per atom required for thermalization. The
product $nv$ depends on both the size and temperature, T, of the
trapped sample. These are measured by observing the expansion of
an equilibrated sample after release from the magnetic trap. The
rate of expansion yields the temperature, while an extrapolation
back to the release time gives the initial sizes. Using the trap
potential calculated from the field coil geometry we have checked
that the measured initial sizes and temperatures are consistent to
within their uncertainties.

The mean number of collisions each atom undergoes, $\alpha$,
during one relaxation time constant was determined from a
numerical simulation of the experiment using classical Monte Carlo
methods \cite{montecarlo}. For a harmonic trapping potential, the
relaxation simulation yields $\alpha_s=2.5$ for s-wave collisions
and $\alpha_p=4.1$ for p-wave collisions.  The ratio
$\alpha_p/\alpha_s$ can also be determined analytically through an
integration over the angular dependence of scattering. This gives
$\alpha_p/\alpha_s=5/3$, consistent with the Monte Carlo results.

The primary results of this paper are shown in Fig.  2.  While
ordinarily one cannot measure higher order partial wave
contributions to the collision cross section directly, the
Fermi-Dirac statistics of $^{40}$K allow us to probe p-wave and
s-wave interactions independently. The p-wave cross section
$\sigma_p$ is determined from measurements using a spin-polarized
sample ($|F=9/2, m_F=9/2>$ atoms), where s-wave collisions are
prohibited by the quantum statistics.  The s-wave cross sections
$\sigma_s$ are determined from data obtained using a mixture of
two spin states, $|9/2, 9/2>$ and $|9/2, 7/2>$. The magnitude of
the p-wave cross section is surprisingly large and we find that
$^{40}$K has a p-wave shape resonance at a collision energy of
roughly 280 $\mu$K.  At temperatures well below the resonant
energy (less than 30 $\mu$K), a fit to $\sigma_p$ vs $T$ gives
$\sigma_p\propto T^{2.0\pm0.3}$. Thus, we have directly observed
the expected threshold behavior $\sigma_p\propto E^{2}$. In
contrast, $\sigma_s$ exhibits little temperature dependence.  With
these very different temperature dependencies, the collision rate
changes by over two orders of magnitude at our lowest temperatures
depending on the spin mixture of the fermionic atom gas.

To explore this effect further we measure the thermalization rate
vs spin polarization at 9 $\mu$K (see Fig.  3). We control the
relative populations of $|9/2,9/2>$ and $|9/2,7/2>$ atoms in a
two-component cloud with a microwave field that drives transitions
to untrapped spin states in the F=7/2 ground state manifold.  The
trap bias magnetic field breaks the degeneracy of the hyperfine
ground-state splitting (1.286 GHz at zero field \cite{splitting})
so that the different spin-states can be removed selectively (see
Fig. 3 inset).  For the data shown in Fig.  3 the fraction of
atoms in the $|9/2,9/2>$ state $f_{m_F=9/2} \equiv
{N_{m_F=9/2}\over N_{m_F=9/2}+N_{m_F=7/2}}$, was varied smoothly
from 70 to 100$\%$ by varying the power of an applied microwave
field (frequency swept) that removes a portion of the $|9/2,7/2>$
atoms.

The thermalization of mixed spin-state samples depends on both
s-wave and p-wave collisions.  The data in Fig.  3 can be fit to a
simple model given by:
\begin{eqnarray}
1/\tau=n_{1,2}({2 \over \alpha_s} \sigma_s + {2 \over \alpha_p}
{\sigma_p \over 2}) v + (n_{1,1} +n_{2,2}) {2 \over \alpha_p}
\sigma_p v, \nonumber
\end{eqnarray}
where $n_{i,j}$ is the density-weighted density between two
species given by $n_{i,j}= {1\over{N_1+N_2}} \int n_i(r) n_j(r)
d^3r$ and the subscripts 1 and 2 stand for the two relevant spin
states.  Since the magnetic moments of the $|9/2,9/2>$ and
$|9/2,7/2>$ atoms are only slightly different we make the
simplifying assumption that these states have identical spatial
profiles in the trap.  A fit using the above model with $\sigma
_s$ and the ratio $\sigma _p / \sigma _s$ as free parameters shows
good agreement with the data in Fig. 3.  In addition to
demonstrating the type of control over collision rates that is
available in a trapped gas of fermionic atoms, this measurement of
$\sigma_p/\sigma_s$ at low temperature provides a sensitive
constraint on the triplet scattering length. The s-wave
cross-sections shown in Fig. 2 were extracted using the above
equation, however at these low temperatures $\sigma_p$ is
relatively small and the measured thermalization rates are due
primarily to s-wave interactions.

To compute the scattering cross sections for comparison with these
data, we first identify singlet and triplet potassium potentials
\cite{triplet} that have been determined spectroscopically. At
large interatomic separation $R$ these potentials are matched
smoothly to the long-range form of Marinescu {\it et al.}
\cite{csix}.  We add an additional correction to these potentials'
inner walls \cite{Burke}, enabling us to tune the scattering
lengths over their entire ranges $-\infty<a<\infty$. We set the
singlet scattering length's value at $a_s=104 a_0$ \cite{Burke}
where $a_0$ is the Bohr radius, but leave the triplet scattering
length $a_t$ as a free parameter to be determined by the
experiment.  We note that the present experiment is relatively
insensitive to the value of $a_s$ since even the
$|9/2,9/2>+|9/2,7/2>$ process is strongly triplet-dominated and no
singlet resonance occurs near threshold; indeed, varying $a_s$
over its range of uncertainty, $101<a_s<107$ \cite{Burke}, does
not change the fit.

After computing cross sections as a function of collision energy,
we determine temperature-dependent cross sections by computing a
thermal average over collision events, weighted by the collision
energy. Using this type of thermal averaging to account for a
temperature-dependent cross section is supported by Monte Carlo
studies \cite{montecarlo}. To make a fit to the data, we compute
$\chi^2$ while floating both $a_t$ and a multiplicative factor
$\epsilon$ which scales simultaneously the computed $\sigma_s(T)$
and $\sigma_p(T)$. This factor is required to accommodate a
$\pm50\%$ systematic uncertainty in the experimental determination
of absolute cross sections (primarily from $N$). Our global best
fit occurs for $a_t=157\pm20 a_0$ and $\epsilon=1.6$, with a
reduced $\chi^2$ of 3.8; the corresponding cross sections are
plotted as lines in Fig. 2.  The uncertainty in $a_t$ reflects a
doubling of the fit $\chi^2$ and includes a $\sim 2 a_0$
uncertainty arising from varying $C_6$ over its range
$3600<C_6<4000$ a.u. \cite{csix}. Our nominal potential gives a
p-wave shape resonance at $\sim280 \mu$K in collision energy, with
an asymmetric lineshape whose FWHM is $\sim400 \mu$K.

    The relatively small uncertainty on the value of $a_t$ is
attributable to the fact that we can simultaneously fit s-wave and
p-wave collision data having little relative uncertainty.  The
value of $a_t$ for $^{40}$K determined here does not agree well
with reference \cite{Cote}, highlighting the importance of low
temperature data in determining accurate potentials.  Our
measurement is however in good agreement with the value
$a_t=194^{+172}_{-42}$ obtained recently from an analysis of
photoassociation spectroscopy of $^{39}K$ \cite{Burke}.   This
agreement between two fundamentally different experiments is very
encouraging, and suggests that the potassium scattering lengths
are now fairly well determined \cite{Jimnote}. Using our new
value, we have tabulated in Table 1 the resulting triplet
scattering lengths for collisions between the different potassium
isotopes.

 One of the important applications of these thermalization rate
measurements is in determining the feasibility of various
evaporative cooling strategies for this fermionic atom gas.  The
large p-wave cross section makes evaporation of a spin-polarized
sample possible, but only at $T \gtrsim 20 \mu K$. To reach lower
temperatures, which is likely to be necessary for producing
quantum degenerate samples, evaporation of a mixed spin-state
sample should work well given the suitably large s-wave elastic
collision cross section.

Indeed, for the data presented in this letter we vary the
temperature of the sample through forced evaporation.  To
facilitate evaporative cooling, we increase the collision rate by
ramping to a $\nu_r=133\pm5$ Hz trap. Evaporation then proceeds
using a mixed spin state sample and applying a microwave field to
remove selectively the most energetic atoms (in both spin states).
With these initial attempts at evaporatively cooling of $^{40}$K
we can lower the sample temperature from 100\ $\mu$K to 5 $\mu$K.
We see runaway evaporation, where the collision rate in the gas
increases as the temperature decreases.  While the samples
discussed in this work are still far from quantum degeneracy, this
initial success and our measurement of a large s-wave elastic
collision cross section bode well for future progress toward this
goal.

This work is supported by the National Institute of Standards and Technology and
the National Science Foundation.  The authors would like to express their
appreciation for useful discussions with C.  Wieman and E.  Cornell.

\epsfxsize=3 truein \epsfbox{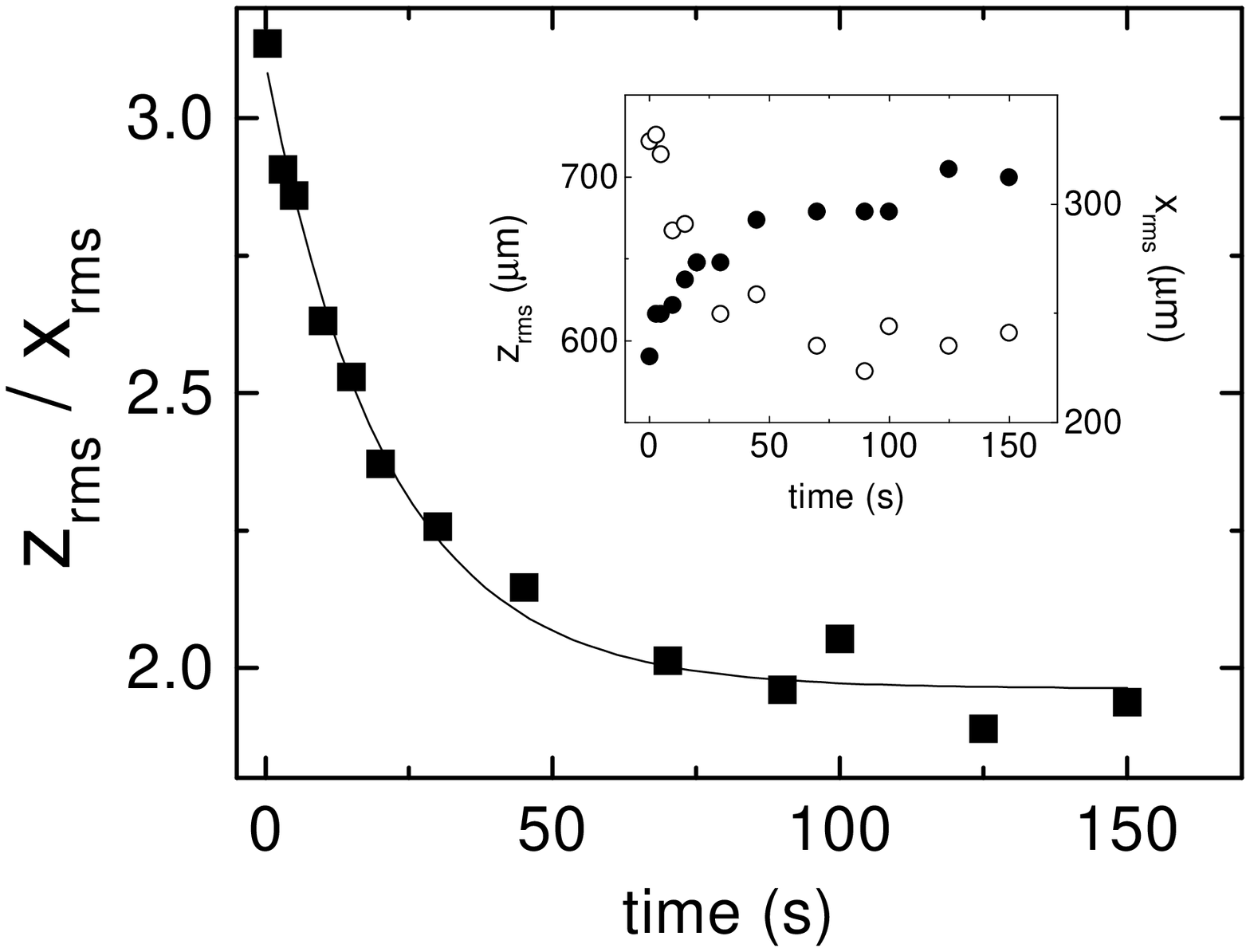}
\begin{figure} \caption{Example of thermalization data.
The inset shows the axial size, $z_{rms}$ ($\bullet$), and the
radial size, $x_{rms}$ ($\circ$), imaged after 2.7 ms of free
expansion, relaxing as the trapped atoms rethermalize via elastic
collisions.  At time=0 the cloud is taken out of equilibrium by
changing $\nu_r$ from 133 to 44Hz.  A fit (line) to the aspect
ratio vs time is used to extract the relaxation rate.
\label{fig1}}
\end{figure}

\epsfxsize=3 truein \epsfbox{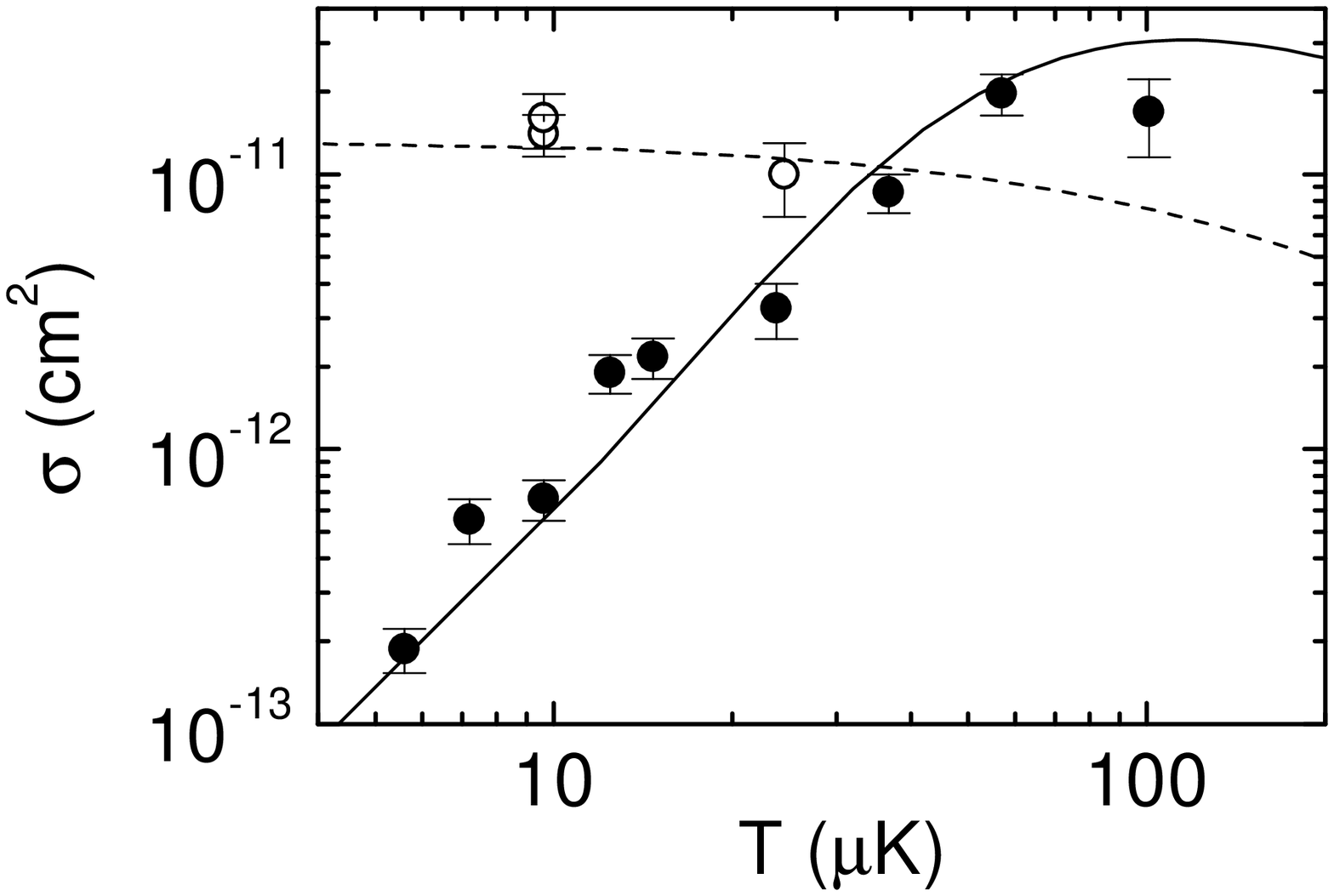}
\begin{figure} \caption{Elastic cross sections vs. temperature.
The s-wave cross section ($\circ$), measured using a mixture of
spin states, shows little temperature dependence.  However, the
p-wave cross section ($\bullet$), measured using spin-polarized
atoms, exhibits the expected threshold behavior and is seen to
vary by over two orders of magnitude.  The lines are a fit to the
data, as described in the text, yielding $a_t=157\pm20 a_0$.
\label{fig2}}
\end{figure}

\epsfxsize=3 truein \epsfbox{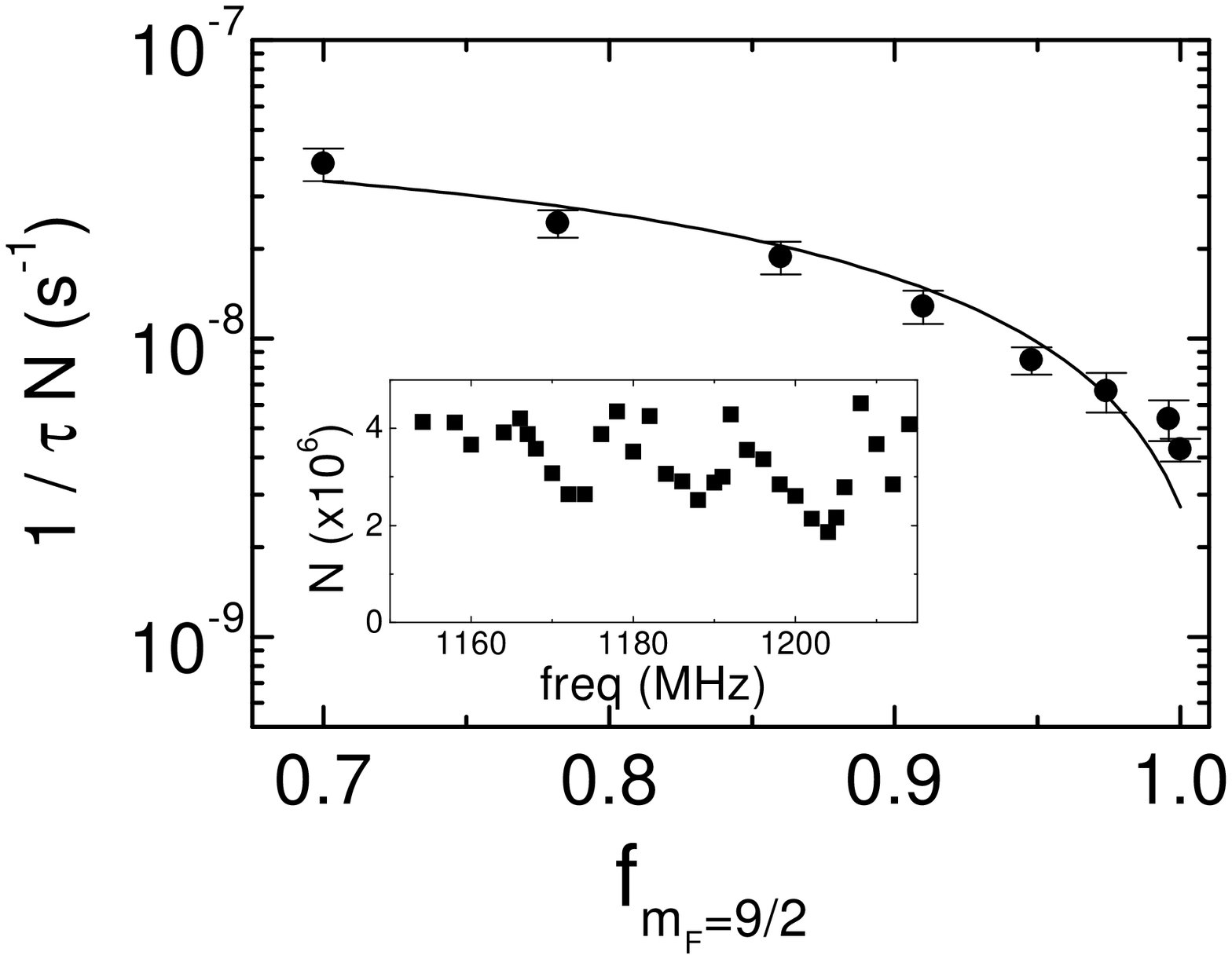}
\begin{figure} \caption{Dependence of $\sigma$ on
spin composition at $T=9 \mu K$.  A quantity proportional to
$\sigma$, $1/\tau N$, is measured as a function of the fraction of
atoms in one of two trapped Zeeman spin states, $f_{m_F=9/2}$. The
inset shows the number of atoms remaining after application of
microwaves at the indicated frequency. The three features
correspond to removal of trapped atoms in particular spin states
and can be used to measure or control the spin composition of the
atom gas. \label{fig3}}
\end{figure}

\begin{table}[tbp]
\caption{Triplet scattering lengths $a_t$ in Bohr radii for
collisions between potassium isotopes.}
\begin{tabular}{lcc}
Isotopes & Nominal $a_t$ & Range \\ \tableline &  &   \\ $40+40$ &
157  & $136 < a_t < 176$\\  $39+39$ & $ -44 $ & $-80 < a_t < -28$
\\  $41+41$ & 57 & $49 < a_t < 62$ \\  \tableline & & \\
 $39+40$ & 3600 & $a_t > 500$ or $a_t < -900$
\\  $39+41$ & 164 & $140 < a_t < 185$ \\  $40+41$
& 93 & $83 < a_t < 99$
\end{tabular}
\end{table}


\begin{references}
\bibitem{SDL} We use SDL power amplifier part SDL-8630-E; SDL is a trade name used for
identification purposes only and does not constitute an
endorsement by the authors or their institutions.
\bibitem{Jimnote} The result of  Ref. \cite{Burke}is
somewhat sensitive to the value of $C_6$, and is in better
agreement with the present result for $C_6=4000$ a.u.
\end{references}
\end{document}